\documentclass[twoside]{article}

\usepackage{graphicx}
\usepackage{fancyhdr}
\usepackage{multicol}
\usepackage{styl-ap}

\begin{document}
\title{Data Analysis of Globular Cluster Harris Catalogue in view of the King models and their dynamical evolution. I. Theoretical model.}
\author{Marco Merafina\work{1}, Daniele Vitantoni\work{1}}
\workplace{Department of Physics, University of Rome La Sapienza, Piazzale 
Aldo Moro 2, I-00185 Rome, Italy}
\mainauthor{marco.merafina@roma1.infn.it}
\maketitle

\begin{abstract} 
We discuss the possibility to analyze the problem of gravothermal 
catastrophe in a new way, by obtaining thermodynamical equations to apply 
to a selfgravitating system. By using the King distribution function in 
the framework of statistical mechanics we treat the globular clusters 
evolution as a sequence of quasi-equilibrium thermodynamical states. 
\end{abstract}

\keywords{Globular Clusters - Gravothermal Catastrophe - King Models - 
Thermodynamical Stability}

\begin{multicols}{2}

\section{Introduction}

Globular clusters (GCs) are stellar systems with masses within the interval
$10^{4}-10^{6}\ M_\odot$, containing a number of stars of the order of 
$10^{5}$. They are considered as nearly spherical systems due to their low 
values of eccentricity $e$; at least 50\% of GCs have $e<0.1$ and there 
are no clusters with $e>0.2$. The core radius $r_{c}$, namely the radial 
coordinate at which the brightness becomes one half of the corresponding 
value at the center of the system, is almost 10 pc, whereas the tidal 
radius $r_{t}$, which is the biggest spatial extension of the cluster 
allowed by the external tidal field, is typically around 50 pc.

For their symmetry and age, there is the possibility to test the evolution 
of a GCs studying a classical single mass King model (King, 1966) in 
relation to thermodynamical instability phenomena. In fact, in the 
analysis of the evolution of GCs, stellar encounters strongly contribute 
in phase space mixing of stellar orbits. In this scenario, thermodynamics 
plays a centrale role in the gravitational equilibrium and stability of 
these clusters, being the average binary relaxation time shorter than their old absolute age which ranges between 10 to 13 Gyr.

This means that Fokker-Planck approximation, which takes into account the nature of collisions in globular clusters, can determine the distribution function relevant for obtaining the equilibrium configurations of these systems, whereas the tidal effects due to the presence of Galactic gravitational potential are responsible of the confination of the cluster.

On the other hand, the observations of the luminosity profiles of different GCs (King, 1962) show similar curves depending only on different values of the star concentration, giving the possibility to fit them by an empirical law and suggesting a unique distribution function for the whole sample of clusters (King, 1966). This effect can be described as a change of the main parameters of the cluster during the dynamical evolution (Horwitz \& Katz, 1977), which maintains the form of the distribution like in a sort of reversible trasformation of a gas in thermodynamic equilibrium, also in accordance to numerical simulations existing in literature which result in keeping unchanged the distribution of velocities of stars for a wide range of values of the concentration during the time evolution driven by the Fokker-Planck equation.

Therefore, the evolution of globular clusters can be studied by considering small thermodynamic transformations which keep constant the functional form of the velocity distribution of stars like in the framework of Boltzmann statistical mechanics. It is important to note that while the equilibrium is given by the form of the distribution which depends on the Fokker-Planck equation and consider the real nature of collisions, thermodynamics plays a role in the tidal effects acting on the confination of the system, due to a two competitive phenomena: one given by stellar encounters which tend to refresh the tail of high velocities in the distribution and one due to evaporation of stars which prevents the formation of it, maintaining the system in a sort of thermodynamical equilibrium with the same distribution function even if in presence of a cutoff in the velocity of the stars.

\section{The effective potential}
The King DF characterizing the energy distribution of stars with the same 
mass $m$ in a model that describes a spherically symmetric system with isotropic velocity distribution may be written as
$$ 
f(\varepsilon)=B\left[{e^{-(\varepsilon +m\varphi)/k\theta} - 
e^{-(\psi +m\varphi)/k\theta}}\right]\;\; {\rm for}\;\; \varepsilon\leq
\psi\ ,
$$
\begin{equation}
f(\varepsilon)=0 \;\; {\rm for}\;\; \varepsilon > \psi\ .
\label{eq1}
\end{equation}
Here $\psi=m\left({\varphi_R -\varphi}\right)$ is the energy cutoff, 
corresponding to the maximum kinetic energy that a star can have at a 
given radial coordinate $r$, while $\varphi$ is the gravitational potential. This energy is sufficient to reach the border of the equilibrium configuration $r=R$, being also the difference between the value of the gravitational potential at the edge of configuration and the same quantity evaluated at a generic distance $r$ from the center. The quantity $\theta$ is the thermodynamic temperature of the system while $B$ is a constant of normalization.

The behavior of equilibrium solutions for King models has been also 
analyzed by Merafina \& Ruffini (1989) by solving the Poisson equilibrium 
equation in Newtonian regime. We can see the presence of a maximum value 
of the total mass $M$ at increasing values of the central gravitational 
potential $W_0$, which denotes the arising of a sort of thermodynamic 
instability at $W_{0}=1.35$ (Fig.\ref{fig1}).

\begin{myfigure}
\centerline{\resizebox{95mm}{!}{\includegraphics{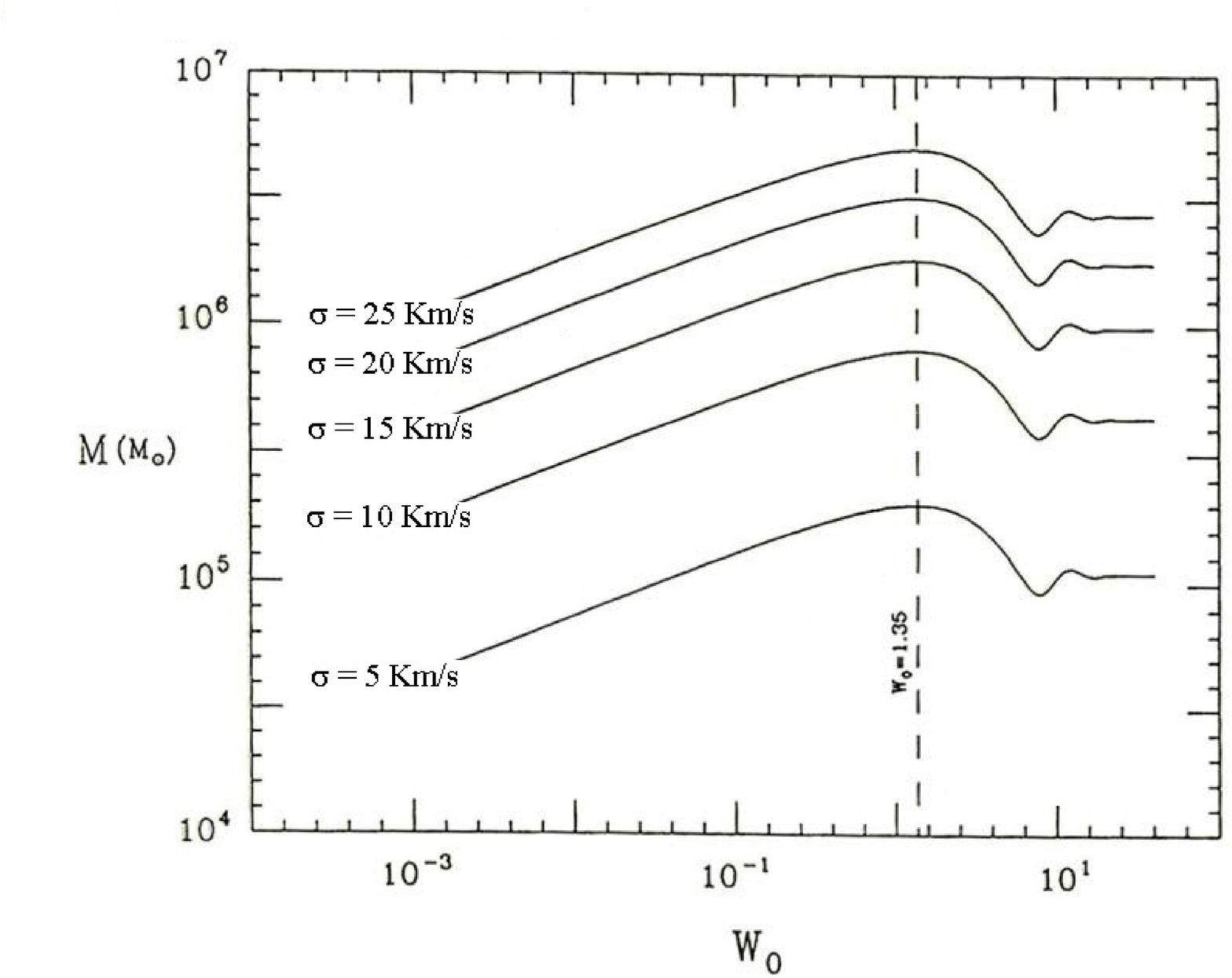}}}
\caption{Mass in function of $W_0$ for families of solutions at different 
values of the velocity dispersion (Merafina \& Ruffini, 1989).}
\label{fig1}
\end{myfigure}

Over this value, we can think the system can evolve towards the loss of thermodynamical equilibrium (\emph{gravothermal catastrophe}), in accordance to the expected evolution of Lynden-Bell \& Wood (1968).

In order to consider thermodynamic transformations in the framework of
statistical mechanics, it is possible to describe the King DF like a 
Maxwell-Boltzmann one by introducing an effective potential. In this way 
the evolution of the King models can be treated as a succession of 
quasi-equilibrium stages by a thermodynamic theory formally equivalent to 
the classical one.

Then, the expression of the effective potential is given by
\begin{equation}
\phi = -k\theta \ln\left[{1-e^{(\varepsilon -\psi )/k\theta}}\right]
\label{eq2}
\end{equation}
and the distribution function can be expressed as
\begin{equation}
f = B e^{-H/k\theta}\ ,
\label{eq3}
\end{equation}
where $H = \varepsilon + m\varphi + \phi$ is the single particle 
Hamiltonian of the system which includes also the gravitational energy of 
the single star. The effective potential is a screen potential which 
restricts the phase space of the available velocities for the stars and takes into account the effect of the tidal forces on the system. In this way, the kinetic temperature $T$ connected with the average velocity of the stars, depending on the radial coordinate $r$, becomes distinguished from the thermodynamic temperature $\theta$, constant all over the equilibrium configuration.

From the modified Boltzmann DF of Eq.\ref{eq3}, we can deduce the 
generalized thermodynamical quantities, as the energy $U$, the 
thermodynamical pressure $\Pi$ and the entropy $S$, related to a shell with radial coordinate $r$. We get 
\begin{equation}
N = AV\int_0^\psi{f\sqrt{\varepsilon}\ d\varepsilon}\ ,
\label{eq4}
\end{equation}

\begin{equation}
U = AV\int_0^\psi{fH\sqrt{\varepsilon}\ d\varepsilon}\ ,
\label{eq5}
\end{equation}

\begin{equation}
\Pi = \frac{1}{3}A\int_0^\psi{f\varepsilon^{3/2}\ 
\frac{dH}{d\varepsilon}\ d\varepsilon}\ ,
\label{eq6}
\end{equation}

\begin{equation}
S = kAV\int_0^\psi{f (1-\ln f)\sqrt{\varepsilon}\ d\varepsilon}\ ,
\label{eq7}
\end{equation}
where we have replaced the costant $B$ with $A$, being 
$B=Ae^{\alpha/k\theta}$ and now $f=Ae^{(\alpha -H)/k\theta}$, while $\alpha = \mu +m\varphi$ is the chemical potential in presence of the gravitational 
potential $\varphi$.

In this way we can rewrite the first law of thermodynamics and obtain a 
new form for the Eulero expression, that include the extensive and 
intensive quantities. We can get also an equation of state formally 
equivalent to classical one which involves the thermodynamical quantities, valid for a shell with radial coordinate $r$. We have
\begin{equation}
dU = \theta dS -\Pi dV +\alpha dN +N\langle dH\rangle\ ,
\label{eq8}
\end{equation}

\begin{equation}
U = \theta S -\Pi V + \alpha N\ ,
\label{eq9}
\end{equation}

\begin{equation}
\Pi V = Nk\theta
\label{eq10}
\end{equation}
and, for kinetic quantities like temperature $T$ and pressure $P$,
\begin{equation}
PV = NkT\ .
\label{eq11}
\end{equation}
Finally, by integrating the expression of $U$ containing the single particle Hamiltonian $H$ (Eq.\ref{eq5}) all over the configuration, we can find an additional term $E_{eff}$ in the expression of the total energy $E_{tot}$ of the system, called \emph{effective energy}
\begin{equation}
E_{tot} = E_{kin} + E_{gr} + E_{eff}\ ,
\label{eq12}
\end{equation}
where $E_{kin}$ and $E_{gr}$ are the total kinetic energy and the total 
gravitational energy, respectively. The partecipation of the effective potential in the total energy corresponds to the account of the tidal potential which determines a finite radius of the cluster. Moreover, Eq.\ref{eq5} defines the energy of the test shell but, for calculating $E_{gr}$, we need to use the expression
\begin{equation}
E_{gr} = \frac{1}{2}\int_0^R \rho\,\varphi\,dV\ .
\label{eq12a}
\end{equation}
 
\subsection{The gravothermal catastrophe}
Thermodynamical instability of a selfgravitating spherical system was 
first studied by Lynden-Bell \& Wood (1968), by considering an isothermal 
sphere (core) confined in a spherical box.  Using the classical form of 
the virial theorem, including a boundary term due to spatial truncation 
of the density prophile, it is possible, for that system, to calculate the critical value of the central gravitational potential $W_{0}=6.55$ after that thermodynamical instability, known as \emph{gravothermal catastrophe}, onsets. It is important to note that such instability takes place only in presence of an external thermal bath exchanging heat with the core and driving the system towards the dynamical collapse.

With the introduction of the effective potential, we can repeat this 
analysis for King models and get another critical value for the 
central gravitational potential $W_{0}=6.9$, which differs from the one obtained by Katz (1980) $W_0=7.4$, due to the additional term in the total energy $E_{tot}$ (see Eq.\ref{eq12}). The most interesting results concern the profile of specific heat for different values of $W_{0}$ (see Merafina et al., in preparation). By analyzing the behavior of the specific heat all over the configuration, we found different results. The expression of the specific heat $C_V=(dQ/d\theta)_V$ arises from Eq.\ref{eq8}, being constant $N$ and $V$, by using the expression
\begin{equation}
dQ = dU - N\langle dH\rangle\ .
\label{eq12b}
\end{equation}

\begin{itemize}
\item For $W_{0}<1.35$, we have equilibrium configurations with positive 
heat capacity all over the system. There are not existing conditions for 
an evolution of the system towards the critical value corresponding to the onset of the gravothermal catastrophe ($W_0=6.9$). Further, this particular value ($W_{0}=1.35$) corresponds to one concerning the first maximum mass we found among the equilibrium solutions (see Fig.\ref{fig1}).

\item For $W_{0}>1.35$, the system shows an external halo with negative 
heat capacity and an internal core with a positive value. The system can evolve by increasing the value of $W_0$ until reaching the critical value in which the gravothermal instability onsets. These evolution can take place
without the necessity of the presence of an external thermal bath, differently from the previously requested condition in the Lynden Bell \& 
Wood model.
\end{itemize}

Results showing the specific heat profiles in function of the radial 
coordinate for different values of $W_{0}$ are summarized in 
Fig.\ref{fig2}.

\begin{myfigure}
\centerline{\resizebox{95mm}{!}{\includegraphics{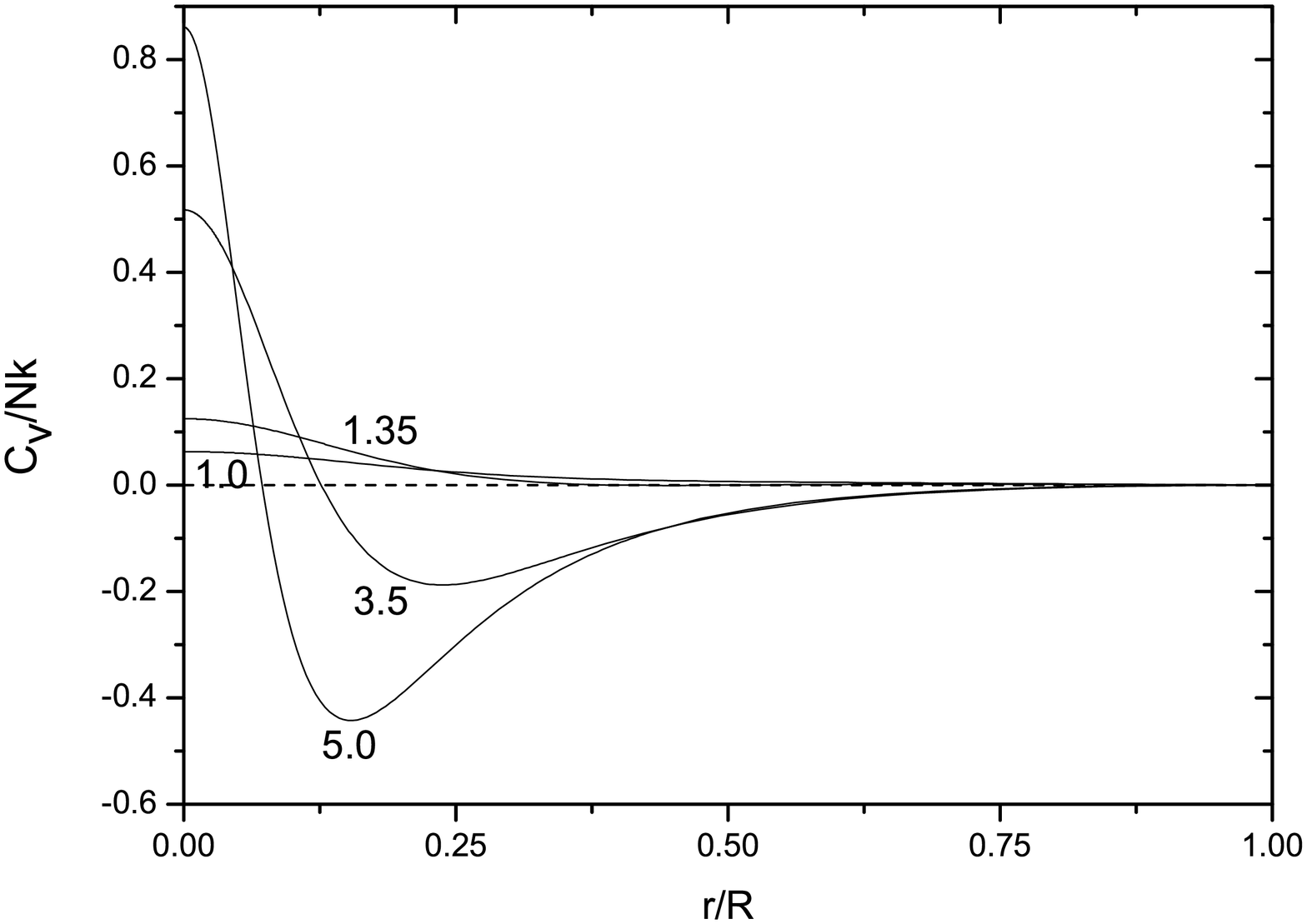}}}
\caption{Behaviour of the specific heat in function of the radial 
coordinate for different values of $W_{0}$.}
\label{fig2}
\end{myfigure}

\subsection{Preliminar observational evidences}
The stability of the King models was analyzed in detail by Katz in 1980, 
with the same investigation carried out by Lynden-Bell \& Wood for the 
isothermal sphere. Katz introduced a new parameter $K$, which corresponds 
essentially to the ratio between the escape velocity and the dispersion 
velocity, both calculated at the center of the cluster. This parameter is 
directly connected with $W_0$. Calculations performed by Katz showed that models become thermodynamically unstable over the value $K=8.1$ ($W_{0}=7.4$). But, analyzing the sample of data coming from Peterson \& King (1975) and Peterson (1976), Katz highlighted an unexplainable gap between the expected value of the sample, $K=7.8$ equivalent to $W_{0}=6.9$, and the one corresponding to the onset of gravothermal instability, resulting at $W_{0}=7.4$ (see Fig.\ref{fig3}).
\begin{myfigure}
\centerline{\resizebox{70mm}{!}{\includegraphics{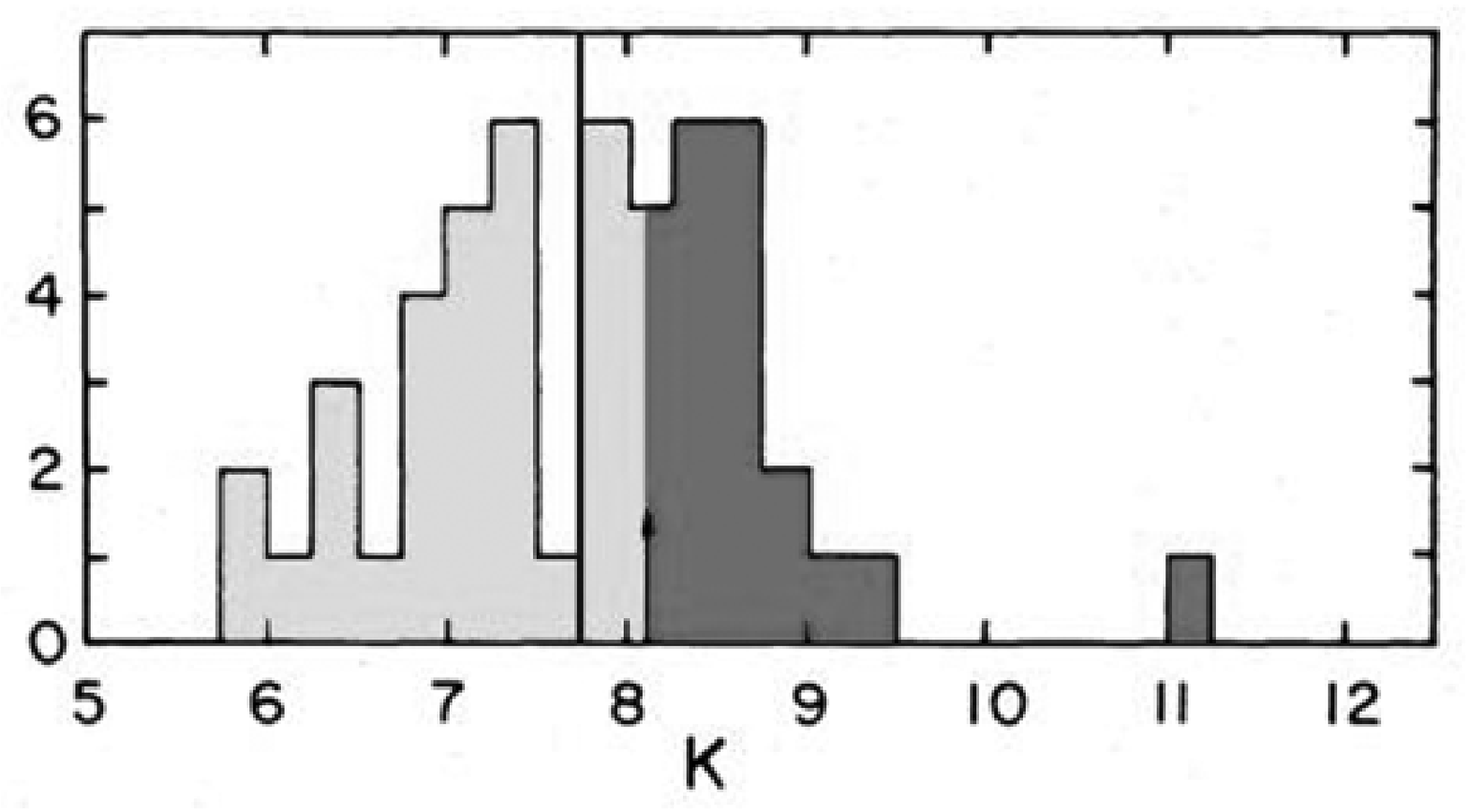}}}
\caption{Distribution of galactic GCs at different values of $K$ (Katz, 
1980).}
\label{fig3}
\end{myfigure}

We indeed expect that GCs had enough time to undergo the gravothermal catastrophe and, therefore, the distribution of GCs in terms of $K$ or $W_{0}$ should peak exactly in correspondence to the critical value. In fact, the primeval Gaussian distribution, approaching the critical value during the evolution, deforms in a non-symmetric Gaussian curve due to the effect of gravothermal catastrophe which progressively subtracts the collapsed GCs with values of $W_0$ larger than the critical value. For these reasons, the resulting distribution must present a maximum which corresponds to the critical value.

It is remarkable to note that, with the introduction of the effective 
potential in the study of the thermodynamical instability, we obtain a 
critical value, $W_{0}=6.9$, that bridges this gap and corresponds 
exactly to the expected value of the sample. This correspondence becomes much more evident by considering the sample of Harris (1996) with 127 clusters (if we exclude the PCC ones), as well becomes more evident the non-symmetric form of the distribution. On the other hand, by making a z-test in order to verify the statistical significance of the gap between the stability limit $W_0=7.4$ expected by Katz and the peak value of the distribution at $W_0=6.9$, it can be shown that these two values are not compatible within a confidence level of 95\%.

\section{Conclusions}

\begin{itemize}

\item The additional (positive) contribution of the effective potential on the total energy, considering also the virial condition $2E_{kin}+E_{gr}=0$, implies that $E_{tot}=-E_{kin}+E_{eff}$. This enables us to construct models in which the core has a positive heat capacity, allowing to assume the possibility of a survival of the system from the gravothermal cathastrophe which could explain the existence of post core-collapsed objects (PCC).

\item The model is selfconsistent and admits regions with positive and 
negative heat capacity which can exchange energy and produce gravothermal 
instability, without the necessity to assume an external bath as in the 
Lynden-Bell \& Wood model.

\item We obtain a new critical value for the onset of gravothermal instability by the presence of the effective potential. This value coincides with the value of $K$ (or, equivalently, to $W_0$) corresponding to the peak of the GCs distribution, removing the unexplainable difference outlined by Katz. This is an observational evidence of the effects due to the presence of the effective potential, confirmed in the analysis of data of more than 150 GCs contained in the last version of catalogue recently published by Harris in 2010 (see also Harris, 1996).

\end{itemize}

Finally, it may be useful to consider some unsolved problems and 
perspectives in order to develop the analysis of thermodynamical 
instabilities of GCs.

\begin{itemize} 
\item The model is not a multimass one and does not take into account the 
effects in the formation of binary stars. At moment this is a preliminary 
model which has to be improved.

\item The new possibility of measuring transverse velocities of the stars 
in GCs opens important perspectives on the knowledge of the distribution 
of the star orbits and their eccentricity, in order to better develop 
N-body simulations in supporting the validity of the model.

\end{itemize}

\end{multicols}
\end{document}